\documentclass[twocolumn,showpacs,preprintnumbers,amsmath,amssymb,prd]{revtex4}
\usepackage{graphicx}

\newcommand{\del}[1]{ \partial_{#1} }

\def\al{{\alpha}}
\def\lam{{\lambda}}

\def\mE{{{\mathcal E}}}

\begin{document}
\title{New Axisymmetric Stationary Solutions 
of Five-dimensional Vacuum Einstein Equations with Asymptotic Flatness}
\author{Takashi Mishima and Hideo Iguchi} 
\affiliation{
Laboratory of Physics,~College of Science and Technology,~
Nihon University,\\ Narashinodai,~Funabashi,~Chiba 274-8501,~Japan
}
\date{\today}
\begin{abstract}
New axisymmetric stationary solutions of the vacuum Einstein equations 
in five-dimensional asymptotically flat spacetimes are obtained 
by using solitonic solution-generating techniques. 
The new solutions are shown to be equivalent to
the four-dimensional multisolitonic solutions
derived from particular class of four-dimensional Weyl solutions 
and to include 
different black rings from those obtained 
by Emparan and Reall.
\end{abstract}      
\maketitle

\section{Introduction}
Inspired by the 
new picture of our Universe including brane world models
and the prediction concerning the production of higher-dimensional 
black holes in future colliders \cite{Giddings:2001bu},
the studies of 
the spacetime structures in higher-dimensional 
General Relativity revealing 
the rich structure have been performed recently with great intensity. 
 For example, several authors examined some qualitative 
features concerning the black hole horizon topologies 
in higher dimensions \cite{ref0}. 
This possibility of 
the variety of horizon topologies gives difficulty 
to the establishment of theorems 
analogous with the powerful uniqueness theorem 
in four dimensions.
Also several exact solutions involving black holes were obtained 
and the richness of the phase structure of black holes has
been discussed. (See \cite{{Elvang:2004ny},{Elvang:2004iz},{Kol:2004ww}}
and references therein.)
Particularly in the five-dimensional case,
several researchers have tried to search new exact solutions
since the remarkable discovery of a rotating black ring solution 
by Emparan and Reall \cite{ref1}. 
For example, the
supersymmetric black rings \cite{ref2} 
and the black ring solutions under the influence of external fields 
\cite{ref3}
are found.
Also a rotating dipole ring solution was studied \cite{Emparan:2004wy}.

Despite these 
discoveries of black ring solutions,
a systematic way of constructing new solutions 
in higher dimensions has not 
been fully developed as for the four-dimensional case, 
particularly for the nonsupersymmetric spacetimes 
with asymptotic flatness. 
In the case of four dimensions 
solution-generating techniques were greatly developed 
and applied to construct new series of axisymmetric stationary solutions extensively 
\cite{ref4}. 
The solutions corresponding to asymptotically flat spacetimes including the famous multi-Kerr solutions by Kramer and Neugebauer \cite{ref6}
were derived systematically, motivated by the discovery of Tomimatsu-Sato solutions \cite{ref5}.


In this article, 
as a first step towards systematic construction of new solutions in higher dimensions
and general understanding of the rich structure of higher-dimensional black objects,
solution-generating techniques similar to those developed in the four-dimensional case
are applied to five-dimensional General Relativity.
(See Ref. \cite{refKK} for the 
Kaluza-Klein compactification.)  
In the following analysis 
we use the fact that one-rotational five-dimensional 
problems can be reduced to four-dimensional ones.
This reduction was considered by several authors in cases
of the Kaluza-Klein theory \cite{{Mazur:1987},{Dereli:1977}} 
and the five-dimensional expression of Jordan-Brans-Dicke theory \cite{Bruckman:1985yk}.

\section{Basic equations}

We consider 
the spacetimes which satisfy the following conditions:
(c1) five dimensions, (c2) asymptotically flat spacetimes, 
(c3) the solutions of 
vacuum Einstein equations, (c4) having three commuting Killing vectors 
including time translational invariance and 
(c5) having a single nonzero angular momentum component. 
Note that, in general, there can be two planes of rotation in 
the five-dimensional spacetime.
Under these conditions, we show that five-dimensional solitonic 
solution-generating problems can be regarded as some four-dimensional problems. 
This means that
we can use the knowledge obtained in the four-dimensional case.
Then we can generate new solutions 
from seed solutions which correspond to known five-dimensional 
spacetimes. 
Here, for simplicity, 
we adopt
the five-dimensional Minkowski spacetime as a seed solution.
As a result, we obtain a new series of solutions which correspond to 
five-dimensional asymptotically flat spacetimes. 
Although the spacetimes found here have singular objects like 
closed timelike curves (CTC) and naked curvature singularities in general, 
we can see that a part of these solutions 
is a new class of black ring solutions whose 
rotational planes are different from those of Emparan and Reall's \cite{ref1}.

Under the conditions (c1) -- (c5), 
we can employ the following Weyl-Papapetrou metric form 
(for example, see the treatment in \cite{refHAR}), 
\begin{eqnarray}
ds^2 &=&-e^{2U_0}(dx^0-\omega d\phi)^2+e^{2U_1}\rho^2(d\phi)^2
       +e^{2U_2}(d\psi)^2 \nonumber \\ 
       &&+e^{2(\gamma+U_1)}\left(d\rho^2+dz^2\right) ,
\label{eq:WPmetric}
\end{eqnarray}
where $U_0$, $U_1$, $U_2$, $\omega$ and $\gamma$ are functions of 
$\rho$ and $z$. 
Then we introduce new functions 
$S:=2U_0+U_2$ and $T:=U_2$ so that 
the metric form (\ref{eq:WPmetric}) is rewritten into 
\begin{eqnarray}
ds^2 &=&e^{-T}\left[
       -e^{S}(dx^0-\omega d\phi)^2
       +e^{T+2U_1}\rho^2(d\phi)^2 \right. \nonumber \\
&&\hskip 0cm \left.
+e^{2(\gamma+U_1)+T}\left(d\rho^2+dz^2\right) \right]
  +e^{2T}(d\psi)^2.
\label{eq:MBmetric}
\end{eqnarray}
Using this metric form
the Einstein equations are reduced to the following set of equations, 
\begin{eqnarray*}
&&{\bf\rm (i)}\quad
\nabla^2T\, =\, 0,   \\
&&{\bf\rm (ii)}\quad
{
\left\{\begin{array}{ll}
& \del{\rho}\gamma_T={\displaystyle
  \frac{3}{4}\,\rho\,
  \left[\,(\del{\rho}T)^2-(\del{z}T)^2\,\right]}\,\ \   \\[3mm]
& \del{z}\gamma_T={\displaystyle 
\frac{3}{2}\,\rho\,
  \left[\,\del{\rho}T\,\del{z}T\,\right],  }
 \end{array}\right. } \\
&&{\bf\rm (iii)}\quad
\nabla^2\mE_S=\frac{2}{\mE_S+{\bar\mE}_S}\,
                    \nabla\mE_S\cdot\nabla\mE_S , \\  
&&{\bf\rm (iv)}\quad
{
\left\{\begin{array}{ll}
& \del{\rho}\gamma_S={\displaystyle
\frac{\rho}{2(\mE_S+{\bar\mE}_S)}\,
  \left(\,\del{\rho}\mE_S\del{\rho}{\bar\mE}_S
  -\del{z}\mE_S\del{z}{\bar\mE}_S\,
\right)}     \\
& \del{z}\gamma_S={\displaystyle
\frac{\rho}{2(\mE_S+{\bar\mE}_S)}\,
  \left(\,\del{\rho}\mE_S\del{z}{\bar\mE}_S
  +\del{\rho}\mE_S\del{z}{\bar\mE}_S\,
  \right)},
\end{array}\right. } \\
&&{\bf\rm (v)}\quad
\left( \del{\rho}\Phi,\,\del{z}\Phi \right)
=\rho^{-1}e^{2S}\left( -\del{z}\omega,\,\del{\rho}\omega \right),  \\
&&{\bf\rm (vi)}\quad 
\gamma=\gamma_S+\gamma_T,   \\
&&{\bf\rm (vii)}\quad 
U_1=-\frac{S+T}{2},
\end{eqnarray*}
where $\Phi$ is defined through the equation (v) and the function 
$\mathcal{E_S}$ is defined by 
$
\,\mE_S:=e^{S}+i\,\Phi\,.
$
The equation (iii) is exactly the same as the Ernst equation in four dimensions \cite{refERNST}, 
so that we can call $\mE_S$ the Ernst potential. 
The most nontrivial task to obtain new metrics is to solve 
the equation (iii) because of its nonlinearity. 
To overcome this difficulty we can however use the methods already 
established in the four-dimensional case. 
Here we use the method similar to the Neugebauer's 
B\"{a}cklund transformation \cite{Neugebauer:1980} 
or the HKX transformation \cite{Hoenselaers:1979mk}, whose essential idea 
is that new solutions are generated by adding solitons to seed spacetimes. 
The applicability of this method to the five-dimensional problem 
is recognized by the following.
The part in the bracket of Eq. (\ref{eq:MBmetric}) corresponds to a
metric of a four-dimensional stationary axisymmetric spacetime with a
``massless scalar field" $T$, where the function $T$ 
is a  solution of the Laplace equation (i).  Then the four-dimensional part is determined 
by a  solution of the Ernst equation (iii).

For the actual analysis in the following, 
we follow the procedure
given by Castejon-Amenedo and Manko \cite{ref7},
in which they 
discussed a deformation of a Kerr black hole 
under the influence of some external gravitational fields.
When a static seed solution $e^{S^{(0)}}$ for (iii) is obtained,
a new Ernst potential can be written in the form
\begin{equation}
{\cal E}_S = e^{S^{(0)}}\frac{x(1+ab)+iy(b-a)-(1-ia)(1-ib)}
                         {x(1+ab)+iy(b-a)+(1-ia)(1-ib)}, \nonumber
\end{equation}
where $x$ and $y$ are the prolate-spheroidal coordinates:
$
\,\rho=\sigma\sqrt{x^2-1}\sqrt{1-y^2},\ z=\sigma xy\,
$
with the ranges $1\le x$ and $-1 \le y \le 1$,
and the functions $a$ and $b$ satisfy the following 
simple first-order differential equations 
\begin{eqnarray}
(x-y)\del{x}a&=&
a\left[(xy-1)\del{x}S^{(0)}+(1-y^2)\del{y}S^{(0)}\right], \nonumber \\
(x-y)\del{y}a&=&
a\left[-(x^2-1)\del{x}S^{(0)}+(xy-1)\del{y}S^{(0)}\right], \nonumber\\
(x+y)\del{x}b&=&
-b\left[(xy+1)\del{x}S^{(0)}+(1-y^2)\del{y}S^{(0)}\right] , \nonumber\\
(x+y)\del{y}b&=&
-b\left[-(x^2-1)\del{x}S^{(0)}+(xy+1)\del{y}S^{(0)}\right]. \nonumber\\
&&    \label{eq:ab}
\end{eqnarray}
The corresponding expressions for the metric functions can be obtained
by using the formulas shown by \cite{ref7}.  

\section{Generation of new solutions}
Here we adopt the following metric form of the five-dimensional 
Minkowski spacetime as a seed solution, 
\begin{widetext}
\begin{eqnarray}
ds^2 &=&
\,-(dx^0)^2+\left(\sqrt{\rho^2+(z+\lam\sigma)^2}-(z+\lam\sigma) \right)d\phi^2
  +\left(\sqrt{\rho^2+(z+\lam\sigma)^2}+(z+\lam\sigma) \right)d\psi^2
   \nonumber \\
 && +\frac{1}{2\,\sqrt{\rho^2+(z+\lam\sigma)^2}}(d\rho^2+dz^2)
 \nonumber\\
 &=& -(dx^0)^2 
    +\sigma\left(\sqrt{(x^2-1)(1-y^2)+(xy+\lambda)^2}
    -(xy+\lam) \right)d\phi^2
    +\sigma\left(\sqrt{(x^2-1)(1-y^2)+(xy+\lam)^2} 
   \right. \nonumber \\&& \left.
    +(xy+\lam) \right)d\psi^2
   +\frac{\sigma(x^2-y^2)}{2\,\sqrt{(x^2-1)(1-y^2)+(xy+\lam)^2}}
     \left[\frac{dx^2}{x^2-1}+\frac{dy^2}{1-y^2}\right],
 \label{eq:metirc_M}
\end{eqnarray}
where $\lam$ and $\sigma$ are arbitrary real constants. 
In this metric the parameter $\lambda$ can be eliminated
by a coordinate transformation. 
Introducing the new coordinates $r$ and $\chi$: 
$$
\rho=\sigma\sqrt{x^2-1}\sqrt{1-y^2}=r\,\chi,\ \ \ z=\sigma xy=\frac{1}{2}(\chi^2-r^2)-\lam\sigma,
$$
the above metric (\ref{eq:metirc_M}) can be transformed into 
a simple form
\end{widetext}
\begin{eqnarray}
ds^2 &=&
\,-(dx^0)^2+(dr^2+r^2d\phi^2)+(d\chi^2+\chi^2d\psi^2). \nonumber
\end{eqnarray}
However the parameter
$\lambda$ acquires a physical meaning 
after the
solution-generating transformation
because this parameterizes the position of the 
gravitational object from the center.

 From Eq. (\ref{eq:metirc_M}), we can derive the seed functions
\begin{eqnarray}
S^{(0)} &=& T^{(0)} =
\frac{1}{2}\ln
      \left[\sqrt{\rho^2+(z+\lam\sigma)^2}+(z+\lam\sigma)\right] 
  \nonumber \\
   &=& \frac{1}{2}\ln
     \left[\sigma \Biglb( \sqrt{(x^2-1)(1-y^2)+(xy+\lambda)^2}
   \right. \nonumber \\ && \left.
             +(xy+\lam)\Bigrb)\right].
\label{eq:seed}
\end{eqnarray}
For the seed function (\ref{eq:seed}) we obtain
the solutions of the differential equations (\ref{eq:ab}) as 
\begin{eqnarray*}
 a &=& \al\,\, \frac{(x-y+1+\lam)
         +\sqrt{x^2+y^2+2\lam xy+(\lam^2-1)}}
              {\ 2\left[(xy+\lam)+\sqrt{x^2+y^2+2\lam xy+(\lam^2-1)}
                                      \right]^{1/2}},                  \\
 b &=& \beta\,\, \frac{\ 2\left[(xy+\lam)+\sqrt{x^2+y^2+2\lam xy+(\lam^2-1)}
                                         \right]^{1/2}}
                {(x+y-1+\lam)+\sqrt{x^2+y^2+2\lam xy+(\lam^2-1)}}\,,
\end{eqnarray*}
where $\alpha$ and $\beta$ are integration constants.

The explicit expression for the corresponding metric is  
\begin{eqnarray} 
ds^2 &=&-\frac{A}{B}\left[dx^0-\left(2\sigma e^{-S^{(0)}}
         \frac{C}{A}+C_1\right) d\phi\right]^2    \nonumber \\
&& +\frac{B}{A}e^{-S^{(0)}-T^{(0)}} \sigma^2(x^2-1)(1-y^2)(d\phi)^2 \nonumber \\
&&   +e^{2T^{(0)}}(d\psi)^2
+ C_2 B
\left(\frac{x-1}{x+1} \right. \nonumber \\ && \left.  
\cdot
      \frac{Y_{-\sigma,-\lam\sigma}}
           {Y_{\sigma,-\lam\sigma}\,[x^2+y^2+2\lam xy+(\lam^2-1)]}\,
\right)^{1/2}  \nonumber \\
&&\times
\left(\frac{dx^2}{x^2-1}+\frac{dy^2}{1-y^2}\right) ,
\label{eq:metric}
\end{eqnarray}
where $Y_{\pm\sigma,-\lam\sigma}$ are given by
\begin{eqnarray*}
Y_{\pm\sigma,-\lam\sigma}
&=&\sigma^2\Bigl[\,(x \mp y)\sqrt{x^2+y^2+2\lam xy+(\lam^2-1)}
  \\ &&
 +x^2 + y^2+(\lam \mp 1)xy \mp (\lam\pm1) \,\Bigr],  \\
\end{eqnarray*}
and $A$, $B$ and $C$ are defined with $a$ and $b$ as
\begin{eqnarray*}
&&A:=(x^2-1)(1+ab)^2-(1-y^2)(b-a)^2 \,, \\
&&B:=[(x+1)+(x-1)ab]^2+[(1+y)a+(1-y)b]^2 \,, \\
&&C:=(x^2-1)(1+ab)[b-a-y(a+b)]  \\
&&\hskip 0.9cm+(1-y^2)(b-a)[1+ab+x(1-ab)]\,.
\end{eqnarray*}
In the following, 
the constants $C_1$ and $C_2$ are fixed as 
\[
C_1=\frac{\,\,2\sigma^{1/2}\,\al\,\,}{1+\al\beta},\ \ \ 
C_2=\frac{\sigma}{2(1+\al\beta)^2},
\]
to assure that the spacetime 
should asymptotically approach the
Minkowski spacetime globally. 
 From the metric (\ref{eq:metric}), 
we can easily see that the sequence of new 
solutions has four independent parameters: $\lambda$, $\sigma$, $\alpha$ and $\beta$. 

\section{Results and Discussion}

We can show that the spacetime of the solution is asymptotically flat. 
If we take the asymptotic limit, $x \rightarrow \infty$,
in the prolate-spheroidal coordinates, the 
metric form (\ref{eq:metric}) approaches the asymptotic form of 
the Minkowski metric 
\begin{eqnarray}
 ds^2 &\sim& -(dx^0)^2+ \sigma x(1-y)d\phi^2 +\sigma x(1+y)d\psi^2 \nonumber \\
  && +\frac{\sigma}{2x}{dx^2}
 +\frac{\sigma x}{2(1-y^2)}{dy^2}.
\end{eqnarray}
Also the asymptotic form of $\mE_{S}$ near the infinity $\tilde{r}=\infty$ becomes 
\begin{eqnarray}
\mE_{S}&=&\tilde{r}\cos\theta\,
\left[\,1\,-\,\frac{\sigma}{\tilde{r}^2}\,\frac{P(\al,\beta,\lam)}
                                       {(1+\alpha\beta)^2}
     \,+\cdots\right] \nonumber   \\
&&  +2\,i\,\sigma^{1/2}\,\left[\,\frac{\alpha}{1+\alpha\beta}
      \,-\,\frac{2\sigma\cos^2\theta}{\tilde{r}^2}\,\frac{Q(\al,\beta,\lam)}
                                                 {(1+\alpha\beta)^3}
    \,\,+\cdots\,\right],\nonumber \\
&&
\end{eqnarray}
where we introduce new coordinates $\tilde{r}$ and $\theta$ by the relations
\begin{eqnarray*}
x=\frac{\tilde{r}^2}{2\sigma}-\lambda, \ y=\cos 2\theta,
\end{eqnarray*}
and
\begin{eqnarray*}
P(\al,\beta,\lam)
   &=& 
       4(1 + \alpha^2 - \alpha^2 \beta^2) ,    \\
Q(\al,\beta,\lam)
   &=& \alpha(2\alpha^2-\lam+3)-2\alpha^2\beta^3 
  -\beta\left[2(2\alpha\beta+1) \right. \nonumber \\ 
 && \left. \times(\alpha^2+1)+(\lam-1)\al^2(\al\beta+2)\right].
\end{eqnarray*}
From the asymptotic behavior, we can compute
the mass parameter $m^2$ and rotational parameter $m^2a_0$:
\begin{equation}
m^2=\sigma \frac{P(\al,\beta,\lam)}{(1+\alpha\beta)^2},\ \ 
m^2a_0=4\sigma^{\frac{3}{2}}\frac{Q(\al,\beta,\lam)}{(1+\alpha\beta)^3}.
\end{equation}

 \begin{figure}
  \includegraphics[scale=0.30]{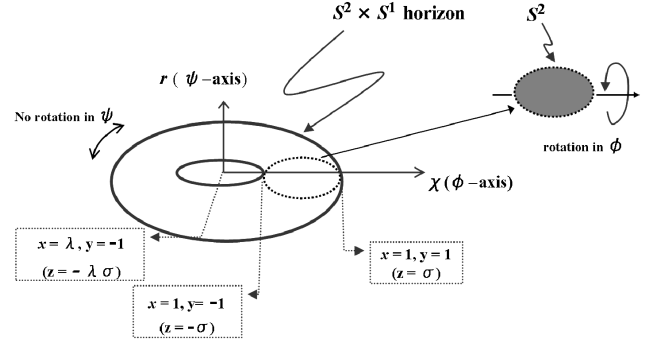}
  \caption{ 
   Schematic diagram of a local ringlike object which resides 
   in the spacetime. Generally some singular behavior appears 
   near the horizon.}
 \end{figure}

The spacetimes generally have some local gravitational objects
which one may regard as black holes. 
Analyzing the rod structure,
which was studied for the higher-dimensional Weyl solutions
by  Emparan and Reall \cite{ref8} and for the nonstatic solutions
by Harmark \cite{refHAR}, 
we can show that
there are event horizons at $x=1$ in these spacetimes. 
In fact, there is a finite timelike rod for $z\in [-\sigma,\sigma]$ with the direction
\begin{equation}
v=(1,\Omega,0), ~~~\Omega=\frac{(1+\alpha\beta)((\lambda+1)\alpha-2\beta)}
             {2\sigma^{\frac{1}{2}}((\lambda+1)\alpha^2 + 2)},
\end{equation}
which corresponds to the region of time translational invariance.
The topology of the event horizon is $S^2 \times S^1$ for $\lambda>1$
as in Figure 1, if it is free of the pathology 
of the Dirac-Misner string \cite{Elvang:2004xi}.
We will discuss this later.

As naturally expected from the presence of the rotation, the new rings
 have ergo-regions.
In fact, the 0-0 component of the metric (\ref{eq:metric})
becomes positive near $x=1$ because the function $A$ becomes
negative there. The limiting form of this componet
at $x=1$ is obtained as
\begin{eqnarray}
 g_{00} &=& \frac{((\lambda+1)\alpha-2\beta)^2(1-y^2)}
{(8(\lambda+y)+(2\beta(1-y)+\alpha(\lambda+1)(1+y))^2}. \nonumber \\
&&
\end{eqnarray}
We show the typical behavior of this componet 
in FIG.2.
 \begin{figure}
  \includegraphics[scale=0.25]{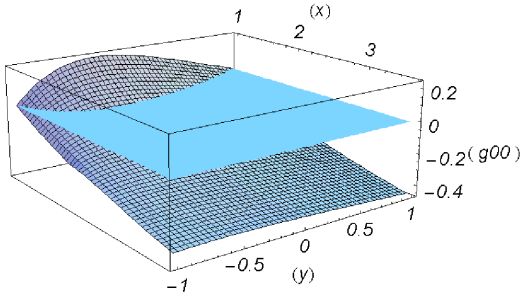}
  \caption{ 
   The behavior of $0$-$0$ component of the metric 
   in the case of $(\alpha,\, \beta,\, \lambda)=(1/2,\, -0.195752,\,2)$.
   The region where the component function is above the level zero 
   corresponds to the ergo-region.}
 \end{figure}


Here we consider the appearance of CTC-regions where the $\phi$-$\phi$ component of the metric becomes negative. At first it can be easily
shown that the value of $g_{\phi\phi}$ is zero at $y=1$. There is no harmful feature around there.
However we can confirm the appearance of CTC from the fact that
the functional form of this component
becomes
\begin{widetext}
\begin{equation}
 g_{\phi\phi}=-\frac{4\sigma(2\alpha \beta^2 + 
        (2 + \alpha^2(\lambda+1))\beta +\alpha(\lambda-1))^2(x^2-1)}
       {(1+\alpha\beta)^2(8\beta^2(\lambda-x)
          +((\lambda-1)(x+1)+ \alpha\beta(\lambda+1)(x-1))^2)},
\label{eq:gpp1}
\end{equation}
\end{widetext}
for the ranges $1<x<\lambda$ at $y=-1$.
This value is always negative except when the parameters satisfy
the following condition
\begin{equation}
2\alpha \beta^2 + 
        (2 + \alpha^2(\lambda+1))\beta +\alpha(\lambda-1)=0.
\label{eq:quad}
\end{equation}
When $\lambda$ and $\alpha$ are given, the parameter $\beta$ should be
\begin{eqnarray}
\beta &=& \beta_{+} \nonumber \\
 &=&
 -\frac{2+\al^2(\lam+1)+\sqrt{\al^4(\lam+1)^2-4\al^2(\lam-3)+4}}{4\al},
\nonumber \\
 &&
\end{eqnarray}
or
\begin{eqnarray}
\beta &=& \beta_{-}  \nonumber \\
 &=&-\frac{2+\al^2(\lam+1)-\sqrt{\al^4(\lam+1)^2-4\al^2(\lam-3)+4}}{4\al}.
  \nonumber \\ &&
\label{eq:noCTC}
\end{eqnarray}
When the parameters satisfy
the condition (\ref{eq:quad}) the solution is free of the pathology 
of the Dirac-Misner string \cite{Elvang:2004xi}. 
Even in this case there can appear the CTC 
when the function $B$ becomes sufficiently small outside the
ergo region.
We can show that the value of $B$ becomes zero at
\begin{equation}
 x=\frac{(\lambda^2-1)\alpha^2-4\beta^2}{4 \alpha\beta}, ~~~ y=0.
\label{eq:zero_point}
\end{equation}
For $\beta=\beta_{+}$, the coordinate value $x$ of (\ref{eq:zero_point}) 
is in its range $x>1$. 
Therefore there appears singular behavior and $g_{\phi\phi}$ becomes
negative in its neighborhood.
While, when $\beta=\beta_{-}$, this singular behavior does not
appear because $x<1$.
As a result, the condition (\ref{eq:noCTC})
makes the singular structure of the spacetimes fairly mild
as seen in FIG. 3, where
the CTC-region which generally appears near the horizon disappears. 
 \begin{figure}
  \includegraphics[scale=0.25]{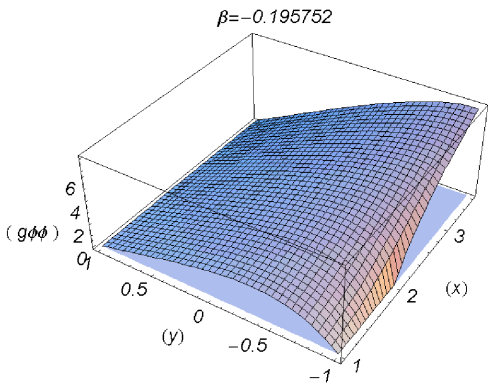}
  \caption{ 
   The behavior of $\phi$-$\phi$ component of the metric 
   in the case of $(\alpha,\, \beta,\, \lambda)=(1/2,\, -0.195752,\,2)$ which 
   satisfies the Eq.(\ref{eq:noCTC}).
   The corresponding component always has non-negative values, while 
   for general case the component becomes negative near the horizon, 
   which means the existence of CTC-regions.}
 \end{figure}


Even for this case, 
there exists a kind of strut structure in this spacetime.
The reason for this is that the effect of rotation
cannot compensate for the gravitational attractive force.
The periods of the coordinates $\psi$ and $\phi$ should be defined as 
\begin{equation*}
 \Delta \psi = 2 \pi \lim_{\chi \rightarrow 0} \sqrt{\frac{\chi^2 g_{\chi\chi}}{g_{\psi\psi}}} 
 ~~~\mbox{and}~~~
 \Delta \phi = 2 \pi \lim_{r \rightarrow 0} \sqrt{\frac{r^2 g_{rr}}{g_{\phi\phi}}},
\end{equation*}
to avoid a conical singularity.
Both $\Delta \psi$ and $\Delta \phi$ for $y=1$, i.e. outside the ring, are $2\pi$. 
While the period of $\phi$ inside the ring becomes 
\begin{equation*}
 \Delta \phi = 2 \pi \frac{\lambda -1 +(\lambda + 1)\alpha \beta}{\sqrt{\lambda^2-1}(1+\alpha\beta)},
\end{equation*}
which is less than $2\pi$ for $1<\lambda<\infty$ with real $\alpha$.
Hence, two-dimensional disklike struts, which appear in the case of 
static rings \cite{ref8}, are needed to prevent the collapse of the rings. 

When $\lam=1$ and $\beta=0$, 
the metric is reduced to the form found 
by Myers and Perry \cite{refMP} which describes a one-rotational spherical 
black hole in five dimensions. 
In fact the metric has the following expression, 
\begin{eqnarray}
ds^2&=&-\frac{p^2x+q^2y-1}{p^2x+q^2y+1}
     \left( dx^0+2\sigma^{1/2}\, \frac{q}{p}\,\frac{1-y}{p^2x+q^2y-1}d\phi \right)^2  \nonumber  \\
   && +\sigma\,\,\frac{p^2x+q^2y+1}{p^2x+q^2y-1}\,
       (x-1)\,(1-y)\, d\phi^2   \nonumber   \\
   && +\sigma(x+1)\,(1+y)\,d\psi^2   \nonumber  \\
   && +{\sigma}\frac{p^2x+q^2y+1}{2p^2}
     \left[ \frac{dx^2}{x^2-1}+\frac{dy^2}{1-y^2} \right], 
\label{eq:MPmetric_1}
\end{eqnarray}
where $p^2=1/(\al^2+1)$ and $q^2=\al^2/(\al^2+1)$.
Introduce new parameters $a_0$ and $m$
through the relations, 
$$
p^2=\frac{4\sigma}{m^2},
    \ \  q^2=\frac{a_0^2}{m^2},
$$
so the metric (\ref{eq:MPmetric_1}) is transformed into
\begin{eqnarray}
ds^2 &=&-(1-\Delta)
    \left[dx^0+\frac{a_0\Delta\sin^2\theta}{1-\Delta}d\phi\right]^2 
    \nonumber \\
&& \hskip -0.3cm\,+\,\frac{1}{1-\Delta}
   \left[\tilde{r}^2+(m^2-a_0^2)\right]\sin^2\theta\, d\phi^2+\tilde{r}^2\cos^2\theta\,d\psi^2 
 \nonumber   \\
&& \hskip -0.3cm\,+\,(\tilde{r}^2+a_0^2\cos^2\theta)
        \left[d\theta^2+\frac{d\tilde{r}^2}{\tilde{r}^2-(m^2-a_0^2)}\right],
\label{eq:MPmetric_2}
\end{eqnarray}
where $\Delta:=m^2/(\tilde{r}^2+a_0^2\cos^2\theta)$.
The line-element (\ref{eq:MPmetric_2}) 
is exactly the same form found by Myers and Perry. 


In some limiting cases with the relation (\ref{eq:noCTC}),  
the corresponding solutions are reduced to the well-known solutions 
like the static black rings or the rotational black strings corresponding to  
(four-dimensional Kerr spacetime) $\times$ {\bf R}. 
The former case is realized when 
we take the limit $\al \rightarrow 0$
and the latter is realized
when the parameter $\lambda$ goes to infinity
under the condition: $\al = \tilde{\al}\times\sqrt{2/\lambda}$
with $-1<\tilde{\al}<1$.


Finally we comment on the four independent parameters. The parameters
$\lambda$ and $\sigma$ characterize the size and mass of 
the local object which resides in the spacetime. 
Appropriate combinations of
$\alpha$ and $\beta$ can be considered as
the Kerr-parameter and the NUT-parameter in four-dimensional case. 
For example,
the five-dimensional analogue of Kerr-NUT solution is obtained 
by setting $\lambda=1$. In this case the parameter
$\beta$ can be considered as a NUT-like parameter 
because, as we have already shown,
the one-rotational Myers-Perry solution
is realized when $\beta=0$.
Also in the case of string like solution, $\lambda \rightarrow
\infty$, the four-dimensional part of this solution corresponds with
the Kerr solution when $\alpha=-\beta$ and with the NUT solution
when  $\alpha=\beta$. 
\section{summary}
In this letter, we generated the new axisymmetric stationary solutions of 
five-dimensional vacuum Einstein equations from the five-dimensional 
Minkowski spacetime as the simplest seed spacetime.
In particular we found a candidate of another branch of 
one-rotational ``black rings". 
Systematic analysis of the new solutions 
will be presented \cite{refIM}. 


In the method presented here we can also adopt other seed 
spacetimes, so that 
we can find some new spacetimes.
However it should be noticed that the method introduced here can not be 
used for the solution-generation of 
black rings with rotation in two independent planes
because of the metric form (\ref{eq:WPmetric}). 
For this purpose other methods may be used.
One of the most powerful methods would be the inverse scattering 
method \cite{refBZ}, 
which was applied to a five-dimensional string theory 
system \cite{Herrera-Aguilar:2003ui}
and 
static five-dimensional cases \cite{Koikawa:2005ia}.

\acknowledgments
We are grateful to Tetsuya Shiromizu for careful reading the manuscript
and useful comments. We also thank Tokuei Sako for his helpful comments 
on the manuscript. 
TM thanks Hermann Nicolai and Stefan Theisen
for useful conversations and the AEI for comfortable hospitality.
This work is partially supported by Grant-in-Aid for Young Scientists (B)
(No. 17740152) from Japanese Ministry of Education, Science,
Sports, and Culture
and by Nihon University Individual Research Grant for
2005.


%
\end{document}